\documentclass[floats,twocolumn,floatfix,amsmath,showpacs,unsortedaddress,superscriptaddress]{revtex4}

\usepackage{graphics}
\usepackage{SIunits}
\usepackage{amsmath}

\newcommand{\ket}[1]{\ensuremath{| #1 \rangle}}

\begin{document}

\title{Coherent Manipulation of Atomic Qubits in Optical Micropotentials}
\author{A. Lengwenus}
\author{J. Kruse}
\affiliation{Institut f\"ur Angewandte Physik,
        Technische Universit\"at Darmstadt, 64289 Darmstadt, Germany}
\author{M. Volk}
\affiliation{Swinburne University of Technology, Melbourne,
Australia}
\author{W. Ertmer}
\affiliation{Institut f\"ur Quantenoptik,
        Universit\"at Hannover, 30167 Hannover, Germany}
\author{G. Birkl}
\email{gerhard.birkl@physik.tu-darmstadt.de} \affiliation{Institut
f\"ur Angewandte Physik, Technische Universit\"at Darmstadt, 64289
Darmstadt, Germany}

\date{\today}


\begin{abstract}
We experimentally demonstrate the coherent manipulation of atomic
states in far-detuned dipole traps and registers of dipole traps
based on two-dimen\-sional arrays of microlenses. By applying
Rabi, Ramsey, and spin-echo techniques, we systematically
investigate the dephasing mechanisms and determine the coherence
time. Simultaneous Ramsey measurements in up to 16 dipole traps
are performed and proves the scalability of our approach. This
represents an important step in the application of scalable
registers of atomic qubits for quantum information processing. In
addition, this system can serve as the basis for novel atomic
clocks making use of the parallel operation of a large number of
individual clocks each remaining separately addressable.
\end{abstract}
\maketitle%
\section{Introduction}\label{intro}
In recent years, there has been growing interest in the
experimental realization of scalable configurations for quantum
information processing \cite{QIP_Overview}. Approaches based on
neutral atoms in microscopic trapping potentials offer a promising
combination of scalability, high decoupling from environmental
sources of decoherence, and advanced techniques for the
manipulation of internal and external degrees of freedom. Trapping
can be achieved either by electric and magnetic potentials created
by micro-fabricated charge- or current-carrying structures
\cite{Atomchips_Overview}, or by optical dipole potentials in the
form of single traps \cite{Grangier1}, interfering laser beams
(optical lattices) \cite{Optical_Lattices} or by light fields
tailored by micro-fabricated optical elements \cite{Birkl1}.

\begin{figure}[b]
    \begin{center}
        \resizebox{0.3\textwidth}{!}{%
        \includegraphics{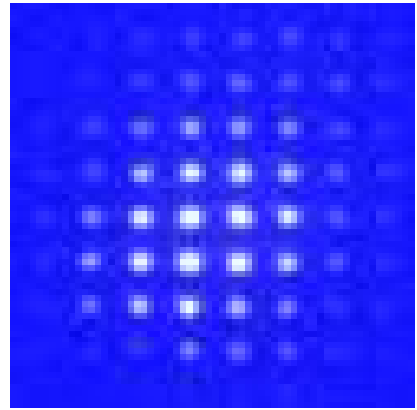}}
    \end{center}
    \vspace{0.5mm}
    \caption{Fluorescence image of about 50 atom samples trapped in a
    two-dimensional array of dipole traps.
    The traps have a waist of \unit{1.7}{\micro\meter} and are separated by \unit{54}{\micro\meter}.
    The central trap contains several 100 $^{85}$Rb atoms.}
    \label{fig:array}       
\end{figure}

In our work, we employ micro-fabricated arrays of diffractive or
refractive lenses to create two-dimensional registers of dipole
traps (Fig. \ref{fig:array}). In the individual traps, we store
neutral rubidium ($^{85}$Rb) atoms in order to generate
two-dimensional registers of quantum bits (qubits) encoded in the
internal or external states of the atoms. We have already
demonstrated the preparation of up to 80 qubit ensembles trapped
in parallel, the individual addressability of each register site,
the controlled preparation of qubit states in each site, and the
independent readout of the qubit state of each site \cite{Dumke1}.

In this paper, we present the next important step towards a
functional quantum processor, namely the coherent manipulation of
qubits encoded in internal atomic states, the investigation of dephasing,
and the determination of the coherence time.
One important feature is the demonstration that
the coherent qubit manipulation and the corresponding readout of the
result of this manipulation can be performed in a scalable fashion
for a large number of sites of the qubit register in parallel.
\section{Experimental setup}\label{qubits}
\begin{figure}[b]
    \begin{center}
        \resizebox{0.40\textwidth}{!}{%
        \includegraphics{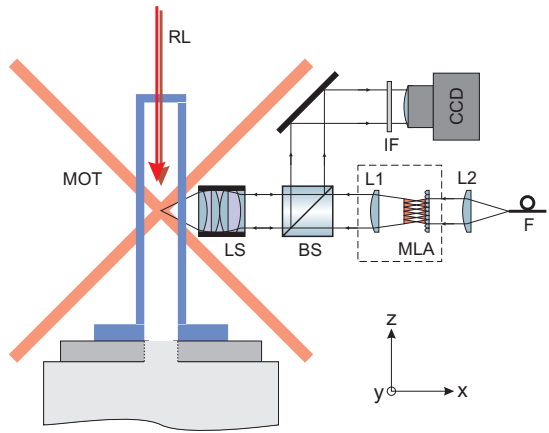}}
    \end{center}
    \vspace{0.5mm}
    \caption{Experimental setup for trapping atoms in arrays of dipole
    traps (not to scale). A microlens array (MLA) is illuminated by the
    light of a TiSa laser delivered by an optical fiber.
    The focal plane of the lens array is
    transferred by lens (L1) and lens system (LS) into the
    MOT-region. Fluorescence of the atoms is detected in the reverse direction:
    behind the lens system (LS) the light is separated by a polarizing beamsplitter
    (BS) and imaged onto a CCD chip. The Raman lasers (RL)
    enter through the end surface of the cell.}
    \label{fig:setup}       
\end{figure}

The experiments are performed with rubidium atoms prepared
by laser cooling and trapping techniques.
The experimental setup is depicted in Fig. \ref{fig:setup}. The
central part is a glass cuvette (\unit{22}{\milli\metre} x
\unit{22}{\milli\metre} x \unit{52}{\milli\metre}) which is
attached to the main vacuum system by a glass-to-metal adaptor.
The pressure is kept below $1\cdot10^{-9}\textrm{ mbar}$ using an
ion-getter pump. We control the partial pressure of rubidium by
either switching on two dispensers in the vacuum chamber, or by
illuminating the glass cell with UV-light at
\unit{395}{\nano\metre} emitted by an array of high intensity
LEDs. While the UV-light is on, atoms which had been adsorbed at the
surface of the glass cell can desorb and defuse into the vacuum,
thereby increasing the partial pressure of rubidium
\cite{Klempt1}.
We capture and precool atoms in a magneto-optical trap (MOT) which
has a magnetic field in anti-Helmholtz configuration with a
gradient of $14 \textrm{ Gauss}/\textrm{cm}$ along the symmetry
axis. Three retroreflected laser beams, which are orthogonal in
orientation and red-detuned by two linewidths relative to the
$5S_{1/2}$, $F=3$ $\rightarrow$ $5P_{3/2}$, $F=4'$ transition,
complete the MOT setup. The total power of the three incoming
cooling beams is about \unit{2}{\milli\watt}, where each beam has
a waist of about \unit{3.5}{\milli\metre}. Repumping light
resonant to the $5S_{1/2}$, $F=2$ $\rightarrow$ $5P_{3/2}$, $F=3'$
transition pumps atoms from the $F=2$ groundstate back into the cooling cycle.
The glass
cell is anti-reflection coated on the outside to maximize
transmission of the cooling and repumping beams of the
magneto-optical trap.
\\
We trap atoms either in arrays of optical dipole traps or - for
systematic studies - in single traps. To create trap arrays, we
employ the setup shown in Fig. \ref{fig:setup}. A microlens array
(see sec. \ref{simultaneous} for details)
is illuminated by a collimated linear polarized Gaussian laser
beam from a titanium sapphire laser (TiSa). The laser light has a
wavelength of \unit{815}{\nano\meter} and is therefore red-detuned
from the D1 and the D2 line
with an effective detuning of $\delta_{\textrm{eff}}/2\pi =
\unit{-13.04}{\tera\hertz}$. The focal plane of the array is
projected into the glass cell by a telescope. The telescope
consists of an achromatic lens ($f=\unit{80}{\milli\metre}$) and a
diffraction limited lens system with a working distance of
\unit{36}{\milli\metre}. This creates arrays of dipole
traps with a waist of \unit{1.7}{\micro\metre} ($1/e^2$ radius)
and a separation of
\unit{54}{\micro\metre}. Experiments on a single trap are either
performed by selecting one trap out of the array or by removing
the microlens array (MLA) and the transfer lens (L1). The latter
case results in a single dipole trap with a waist of
\unit{9.7}{\micro\meter}. For detection, we illuminate the atoms
with the cooling and repumping beams for
\unit{300}{\micro\second}. The fluorescence light is collected
with lens system LS and then reflected by a polarizing
beamsplitter onto an electron multiplying charge coupled device
camera (EMCCD). The advantage of this type of camera is the
electron multiplication directly on the CCD chip which is
comparable to the performance of an avalanche photodiode. Hence,
readout noise is minimized, which enables us to detect even single
photons. An interference filter (IF) placed in front of the camera
blocks straylight from the TiSa, while transmitting fluorescence
light at \unit{780}{\nano\metre}.

\section{Experimental sequence} \label{experiement} A typical
experimental sequence is as follows: first the MOT is loaded from
the background gas by switching on the magnetic field and the
cooling and repumping beams. After the MOT is loaded, the atoms
are further cooled for \unit{5}{\milli\second} by polarization
gradient cooling. During this period, the TiSa beam is switched on
by an acousto-optical modulator, and atoms are loaded into the
dipole trap, which is superimposed on the MOT and the optical
molasses. Switching off the MOT beams and waiting for another
\unit{50}{\milli\second} ensures that all atoms not trapped in the
dipole trap have left the detection region. The number of loaded
atoms is on the order of 1000 atoms per trap. The temperature is
measured by a time-of-flight method and is approximately
$\unit{40}{\micro\kelvin}$.
\begin{figure}[t]
    \resizebox{0.48\textwidth}{!}{%
    \includegraphics{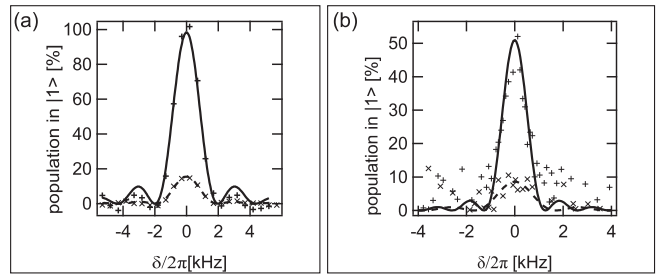}}
    \caption{Efficiency of optical pumping: population in state $\ket{1}$
    normalized to all atoms for (a) free expanding atoms
    and (b) atoms in the central trap of a dipole trap array. Fits to the central resonance (lines)
    are given for data (crosses) with (straight line) and without (dashed line) optical pumping.
    The functional form of the fits is given by the Fourier spectrum of
    a rectangular pulse used during detection (see section \ref{mani} for details).}
    \label{fig:Pumping}       
\end{figure}
\subsection{Qubit preparation and state selective detection}
\label{prepdet} After the atoms have been loaded into the dipole
trap, they have to be initialized for coherent manipulation. We
have chosen the hyperfine states $\ket{0} \equiv \ket{F=2,
m_{F}=0}$ and $\ket{1} \equiv \ket{F=3, m_{F}=0}$ as qubit states,
because of their insensitivity to fluctuations of the magnetic
field. This specific choice is only possible in dipole traps not
relying on the Zeeman shift for atom trapping. As an important
consequence, we can coherently manipulate the qubit states by
driving the in first-order magnetic field insensitive clock
transition $\ket{F=2, m_{F}=0} \rightarrow \ket{F=3, m_{F}=0}$.
However, we have to use a more elaborate optical pumping scheme to
transfer the atoms into the \ket{F=3, m_{F}=0} state for
initialization, after being equally distributed over all seven
$m_{F}$-states at the end of the loading phase from the MOT. For
the pumping process we first switch off the cooling laser beams
and apply a magnetic offset field of \unit{50}{\micro\tesla} (0.5
Gauss) along the z-axis to define a quantization axis. The pumping
is induced by a $\pi$-polarized laser beam, which is resonant to
the $\ket{5S_{1/2}, F=3} \rightarrow \ket{5P_{3/2}, F'=3}$
transition and the MOT repumping light to prevent pumping into the
\ket{F=2} state. For $\pi$-polarized light the \ket{F=3,m_{F}=0}
state is a dark state, so that atoms which end up in this state
stay there during the pumping process. The efficiency of this
pumping process is shown in Fig. \ref{fig:Pumping} for atoms in
the central trap of a dipole trap array with a trap depth of
$k_{B} \cdot \unit{1.0}{\milli\kelvin}$ as well as for free
expanding atoms (i.e. released from the MOT). We observe a pumping
efficiency of up to $100\%$ for free expanding atoms and an
efficiency of $51\%$ for atoms in the dipole potential with $49\%$
of the atoms distributed over the $\ket{F=3, m_{F} \neq 0}$
states. We assume that the reduced efficiency in the dipole trap
results from two photon transitions coupling the
various \ket{m_{F}}-states or from the Stark shift inside the dipole traps.\\
State selective detection is performed by using an additional
intense laser pulse resonant to the $\ket{F=3} \rightarrow
\ket{F'=4}$ transition. If the intensity is chosen to be
sufficiently high ($I/I_{0} \approx 100$), the radiation pressure
force is stronger than the dipole force. The atoms in \ket{F=3}
are then \textit{pushed} out of the dipole trap so that only the
atoms in \ket{F=2} remain.
One has to ensure that the irradiation time is short enough ($<
\unit{300}{\micro\second}$), so that spontaneous decay into the
\ket{F=2} state can be neglected. The atoms in \ket{F=2} are
detected by resonant excitation with the MOT and repumping light
and collecting the fluorescence light with the CCD camera. Note
that this detection mechanism is not only state selective, but
also resolves the atoms in space so that many traps can be
detected at the same time. In combination with the spatial
selective addressability demonstrated in \cite{Dumke1}, this
presents a powerful scheme for preparation and readout of qubit
states for scalable quantum information processing.

\subsection{Coherent manipulation}
\label{mani}%
\begin{figure}[b]
    \resizebox{0.48\textwidth}{!}{%
    \includegraphics{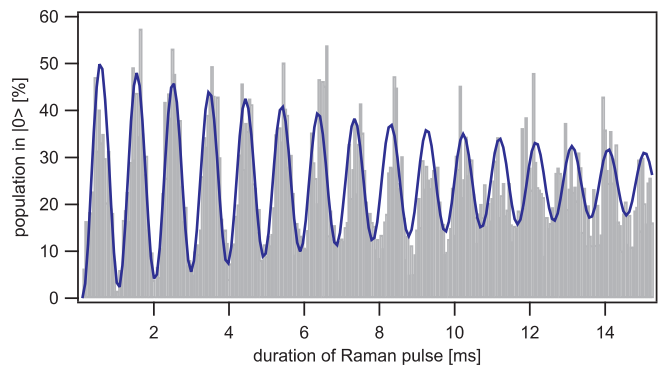}}
    \caption{Rabi oscillations in a dipole trap: population of the
    state $\ket{0}$ normalized to all atoms in the trap before optical
    pumping as a function of the
    duration of the Raman pulse. $98\%$ of the atoms prepared
    in state $\ket{1}$ participate in the first Rabi cycle.}
    \label{fig:Rabi}       
\end{figure}

We use a Raman laser system to coherently couple states \ket{0} and
\ket{1}. The Raman laser system consists of two phase
locked extended cavity diode lasers with a frequency difference
equal or close to the hyperfine splitting of the $5S_{1/2}$ state
(\unit{3.04}{\giga\hertz}). The output radiation can be switched
by an acousto-optical modulator. If we illuminate atoms which are
prepared in the upper qubit state $\ket{1}$ with Raman pulses of
variable length, periodic transfer of the population to and from
the lower qubit state $\ket{0}$ occurs. Here, we make use of the
fact that the magnetic offset field shifts transitions between
other magnetic substates out of resonance. Therefore, atoms
not pumped into $\ket{1}$ during preparation do not
contribute to the subsequent population transfer. Rabi oscillations of
atoms in one of the central traps of a dipole trap array are shown
in Fig. \ref{fig:Rabi}. The trap depth is $k_{B} \cdot
\unit{1.2}{\milli\kelvin}$. Each data point represents a separate
Rabi experiment with differing duration of the Raman pulse.
Measured populations are normalized to the number of trapped atoms
before optical pumping.
Fluctuations in the measured population are mostly of a
statistical nature and occur mainly due to fluctuations of the
number of loaded atoms and imperfect efficiency in preparation and
detection. We fit the solution of the numerically integrated Bloch
equations to the data by taking into account that due to
spontaneous scattering of photons from the trapping beam the
population in the initially empty state \ket{0} increases with
time. We find that $50\%$ of the trapped atoms participate in the
Rabi oscillations. Including the pumping efficiency of $51\%$ (see
Fig. \ref{fig:Pumping} (b)), we find that $98\%$ of the atoms
prepared in \ket{1} contribute to the Rabi oscillations.
From the fit we determine a Rabi frequency of $\Omega_{R}=2\pi
\times \unit{(995 \pm 5)}{\second^{-1}}$. The damping of the
oscillations is caused by dephasing and decoherence effects. Both
effects are indistinguishable in this measurement. We will focus
on dephasing and decoherence effects in the following section.
Maximum coherent population transfer can be achieved by
irradiating the atoms with a resonant Raman pulse of length
$\pi/\Omega_{R} =\unit{(503\pm3)}{\micro\second}$ ($\pi$-pulse).
Such a pulse is employed to pump atoms from \ket{1} to \ket{0}
with high efficiency during the state-selective detection
sequence. Varying the frequency difference of the Raman laser
fields gives the frequency dependent transfer efficiency shown in
Fig. \ref{fig:Pumping}.

\section{Dephasing and coherence} \label{coherence}
A real quantum mechanical system
cannot be perfectly isolated from the environment. This will cause
an initial pure quantum state to evolve into a statistical mixture
of states which manifests itself in two effects, namely dephasing
and decoherence. These effects are responsible for information loss
in a quantum system. Long coherence times are therefore a crucial
parameter for the applicability of a physical system
for quantum information processing. We follow the previously
discussed methods to study decoherence of quantum states of neutral
atoms in optical dipole traps \cite{Davidson1,Ozeri1,Kuhr1}.
A theoretical description of
these effects can be based on the Bloch
equations \cite{Bloch1}:
\begin{subequations} \label{blocheq}
\begin{eqnarray}
    \dot{u} &=& -\delta \cdot v - \frac{u}{T_{2}},\\
    \dot{v} &=& \delta \cdot u - \Omega_{R} \cdot w -
    \frac{v}{T_{2}},\\
    \dot{w} &=& \Omega_{R} \cdot v - \frac{w - w_{\textrm{eq}}}{T_{1}}.
\end{eqnarray}
\end{subequations}
The notation is taken from nuclear magnetic resonance (NMR) physics.
$T_{1}$ is called longitudinal relaxation time and describes the
time constant in which the inversion $w$ of the system evolves to
its equilibrium value $w_{\textrm{eq}}$. This can be caused by
scattering of photons from the trapping laser or by collisions and
can be equated to the coherence time. The transversal relaxation
time $T_{2}$ describes the change in polarization $u$ and $v$ of the
system. It is composed of an irreversible polarization decay time $T_{2}'$
(homogeneous dephasing) and a reversible part $T_{2}^{*}$
(inhomogeneous dephasing):
\begin{equation}
\frac{1}{T_{2}} = \frac{1}{T_{2}'} + \frac{1}{T_{2}^{*}}
\end{equation}
\begin{figure}[t]
    \resizebox{0.48\textwidth}{!}{%
    \includegraphics{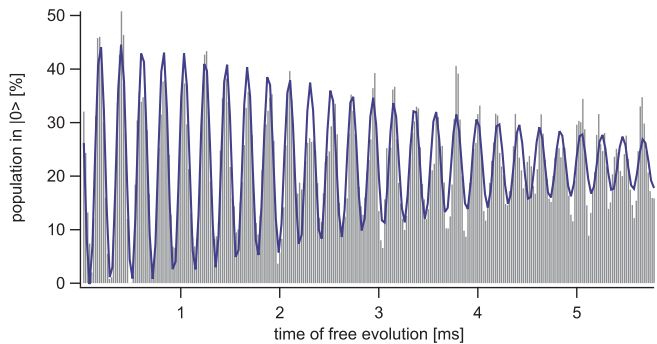}}
    \caption{Ramsey measurement in a single dipole trap. From the fit we get a dephasing time
    $T_{2}^{*}=\unit{(4.08\pm0.22)}{\milli\second}$ and a precession frequency
    of $\delta= 2\pi \times \unit{(4814 \pm 5)}{\hertz}$.}
    \label{fig:ramsey_single}       
\end{figure}
\subsection{Ramsey measurements}
\label{ramsey} In the Rabi experiment, oscillations are driven
with the Raman lasers continuously illuminating the atoms during
the period of coherent manipulation. This can cause scattering of
photons from the Raman light fields as well as a differential
light shift of the qubit states. While the first effect will
destroy coherence the second effect shifts the transition
frequency, which affects the inhomogeneous dephasing time
$T_{2}^{*}$. One can overcome these problems by extending the Rabi
experiment to a Ramsey experiment \cite{Ramsey1}. We first prepare
the atoms in the state $\ket{1}$ as before. A $\pi/2$-pulse
transfers them to a coherent superposition of the two qubit states
\ket{1} and \ket{0}. In the framework of the Bloch model
(\ref{blocheq}), the Bloch vector is flipped into the equatorial
plane ($uv$-plane). Now, the Raman pulses are switched off, which
results in a free precession of the Bloch vector around the
$w$-axis. The angular frequency $\delta = \omega_{\textrm{Atom}} -
\omega_{\textrm{RL}}$ is given by the frequency difference between
the present atomic resonance (generally shifted from the nominal
hyperfine splitting of the $5S_{1/2}$ state with
$\omega_{\textit{HFS}} = \unit{3.0357}{\giga\hertz}$) and the
frequency difference of the Raman lasers. For \textit{resonant}
pulses ($\omega_{\textrm{RL}} = \omega_{\textrm{HFS}}$), $\delta$
is mainly given by the differential light shift induced by the
dipole trap and the quadratic Zeeman shift by the magnetic offset
field. After free precession during time $t$, a second $\pi/2$
pulse is applied which rotates the Bloch vector by 90 degrees
around the $u$-axis. Finally, detection gives the population in
the lower qubit state \ket{0}. Without any dephasing or
decoherence, this results in an oscillation of the probability
$P_{\ket{0}}$ of finding the atom in the state \ket{0} depending
on the free precession time:
\begin{equation}
P_{\ket{0}} = \frac{1-w}{2} = \frac{1}{2} \left( 1+\cos{\delta t}
\right).
\end{equation}
Taking into account that we are working with an ensemble of atoms
which are thermally distributed in the dipole trap one can follow
the calculations in \cite{Kuhr1} giving
\begin{equation}\label{pop_ramsey}
P_{\ket{0}}(t) = A \cdot \alpha(t,T_{2}^{*}) \cos \left[ \delta t
+ \kappa(t,T_{2}^{*}) + \Phi \right] + C
\end{equation}
where
\begin{subequations}
\begin{eqnarray}
\alpha(t,T_{2}^{*}) &=& \left[ 1 + 0.95 \left( \frac{t}{T_{2}^{*}}
\right)^{2} \right]^{-3/2},\\
\nonumber\\
\kappa(t,T_{2}^{*}) &=& -3 \arctan{ \left( 0.97
\frac{t}{T_{2}^{*}} \right)}.
\end{eqnarray}
\end{subequations}
If we assume that the thermal distribution of the atoms is the
dominating cause of inhomogeneous dephasing, one gets a
relation between the dephasing time $T_{2}^{*}$ and the mean
temperature $T$ of the atomic ensemble
\begin{equation}\label{temperature}
T = 1.94 \cdot \frac{\hbar \cdot
\delta_{\textit{eff}}}{k_{\textit{B}} \cdot \omega_{\textit{HFS}}
\cdot T_{2}^{*}}.
\end{equation}
Because of its specific kinetic energy, every atom in a thermal
ensemble experiences a different shift $\delta$ in the dipole trap
due to the fact that for the lower qubit state \ket{0} the
detuning of the dipole trap is larger by $\omega_{\textrm{HFS}}$
than for the upper qubit state \ket{1} (differential light shift).
For example, atoms with lower temperature will be found more at
the bottom of the trap and will therefore experience a larger
differential light shift. In the Bloch model, this is equivalent
to a different precession frequency of the Bloch vectors around
the $w$-axis. If an ensemble of atoms is considered, the Bloch
vectors of all atoms will spread out leading to a reduced
visibility of the Ramsey signal. Equation (\ref{pop_ramsey}) also
embodies the constant offset $A$ and $C$, which are relevant for
fitting experimental data, because they account for the actual
population in the upper and lower qubit states. The phase offset
$\Phi$ accounts for an additional phase shift the system gains
since the Bloch vector not only precesses during the free
evolution time $t$ but
also during the $\pi/2$-pulses.\\
\begin{figure}[t]
    \resizebox{0.48\textwidth}{!}{%
    \includegraphics{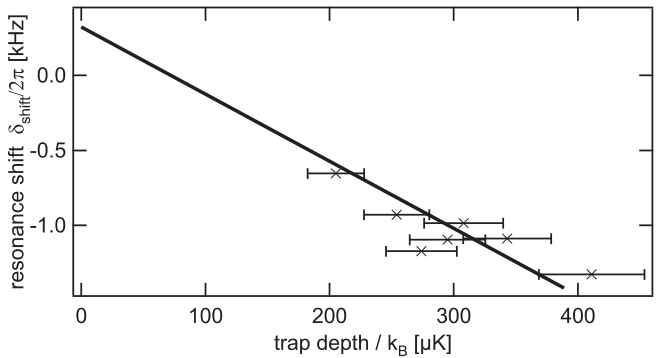}}
    \caption{Shift $\delta_{\textrm{shift}}$ of the atomic resonance as a function of
    trap depth. Experimental data from Ramsey measurements are compared
    to theory (line) which takes into account the differential light shift and the quadratic
    Zeeman effect.}
    \label{fig:ramsey_depth}       
\end{figure}
We investigated Ramsey oscillations in the single dipole trap
setup. The results are depicted in Fig. \ref{fig:ramsey_single}.
Clearly observable are Ramsey oscillations with precession
frequency $\delta$ and a reduction of the modulation amplitude of
the signal with increasing $t$. The precession frequency is
determined by $\delta = \delta_{\textrm{RL}} +
\delta_{\textrm{shift}}$ where $\delta_{\textrm{RL}} =
\omega_{\textrm{RL}} - \omega_{\textrm{HFS}}$ is the detuning of
the Raman lasers from the frequency of the hyperfine splitting.
Additional shifts acting on the atomic states are summarized in
$\delta_{\textrm{shift}}$ which are mainly the differential light
shift and the quadratic Zeeman shift. Due to the high
spectroscopic resolution of the Ramsey technique, we can directly
measure
$\delta_{\textrm{shift}}$ as a function of external parameters.\\
Figure \ref{fig:ramsey_depth} visualizes the measured resonance
shift $\delta_{\textrm{shift}}$ as a function of trap depth.
Typical trap depths range from
$k_{B}\cdot\unit{200}{\micro\kelvin}$ to
$k_{B}\cdot\unit{400}{\micro\kelvin}$.
We calculate the frequency shift using a Raman laser detuning of
$\delta_{\textrm{RL}}=0$ and a quadratic Zeeman shift, which is
induced by the magnetic offset field of \unit{50}{\micro\tesla}
(0.5 Gauss). This gives a frequency offset of
$\delta_{\textrm{shift}}(0) \approx +\unit{320}{\hertz}$ in the
absence of the dipole trap. The solid line in Fig.
\ref{fig:ramsey_depth} represents the calculated shift which
agrees well with our data. From this and similar measurements, we
know that we can determine frequency shifts for atoms trapped in
the dipole potential with a spectroscopic resolution of better
than \unit{100}{\hertz}.

\subsection{Echo measurements}
\label{echo}
\begin{figure}[b]
    \resizebox{0.48\textwidth}{!}{%
    \includegraphics{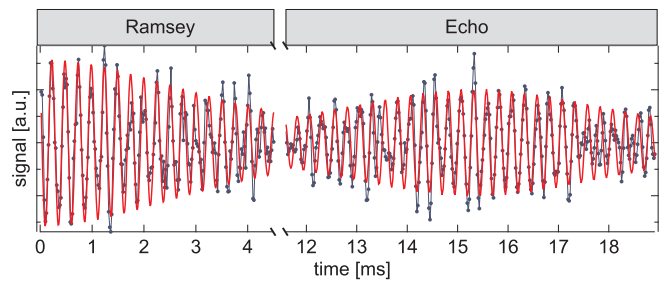}}
    \caption{Ramsey and echo measurement in a dipole trap together
    with a theoretical fit. The $\pi$-pulse which
    reverses the dephasing is applied at $t_1=\unit{7.5}{\milli\second}$. After $2\cdot t_1=\unit{15}{\milli\second}$
    the modulation of the signal is again maximized.}
    \label{fig:Echo}       
\end{figure}
\begin{figure}[t]
    \resizebox{0.48\textwidth}{!}{%
    \includegraphics{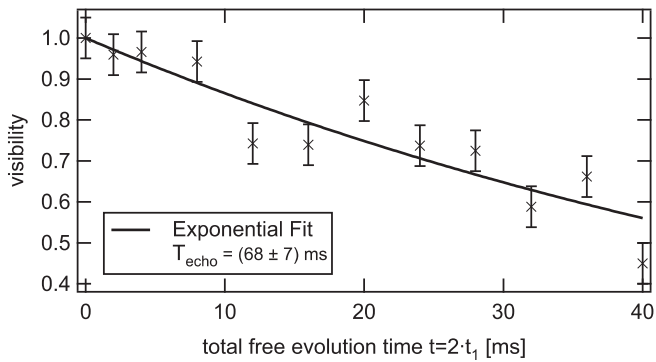}}
    \caption{Visibility of the echo signal at $t=2\cdot t_{1}$.
    Assuming that decoherence caused by spontaneous scattering is the main
    cause for loss of visibility, the data can be fitted by an exponential decay
    (line).}
    \label{fig:Coherence}       
\end{figure}

It is not possible to determine the coherence time based on Ramsey
measurements. As already explained, the decay of the signal can
also be caused by dephasing, so that decoherence effects are not
distinguishable from dephasing effects. In order to check whether
the dominating dephasing mechanism is reversible and to determine
the coherence time of the system, one can extend the Ramsey
measurements to (spin-)echo-spectroscopy \cite{Hahn1,Kurnit1}.
Echo measurements with atoms in optical dipole potentials have
already been presented in \cite{Buchkremer1,Andersen1}. The idea
is to reverse the dephasing which occurs after the first
$\pi/2$-pulse at $t_{0}=0$ by applying an additional $\pi$-pulse
at $t_{1}>t_{0}$. This will lead to a new maximum in the
modulation amplitude at time $t=2 \cdot t_{1}$. A typical echo
measurement together with the corresponding Ramsey measurement and
theoretical fits is shown in Fig. \ref{fig:Echo} for a time
$t_{1}=\unit{7.5}{\milli\second}$. This measurement was performed
in the single trap setup. As expected, the signal modulation is
again maximized at a time corresponding to the echo signal with a
modulation amplitude significantly larger than for the Ramsey
experiment after the same time delay. This is proof that the
reduction of the Ramsey fringes is dominated by dephasing and not
by decoherence. The visibility of the echo signal, which is
defined as the amplitude of modulation of the echo signal at $t=2
\cdot t_{1}$ relative to the amplitude of modulation of the Ramsey
signal at $t=0$, gives information about decoherence and
homogeneous dephasing. The visibility as a function of increasing
$t_{1}$ is shown in Fig. \ref{fig:Coherence}. A reduction of the
visibility being due to decoherence caused by spontaneous
scattering of photons should lead to an exponential visibility
decay with a time constant given by the inverse photon scattering
rate. An exponential fit to the data in Fig. \ref{fig:Coherence}
results in a decay time of
$T_{\textrm{echo}}=\unit{(68.0\pm7)}{\milli\second}$.
%
%
This value agrees well with the inverse spontaneous scattering
rate from the trapping laser light which is
$\Gamma^{-1}_{\textrm{sc}} =\unit{(68.0\pm5)}{\milli\second}$.
From this we draw the conclusion that the reduction of the echo
visibility is a result of spontaneous scattering of photons from
the trapping light. Thus, the echo measurement directly gives the
coherence time $T_{1} \approx T_{\textrm{echo}}$. Although
homogeneous dephasing might be present, the time constant for
$T_{2}'$ should be significantly larger than the time constant for
decoherence $T_{1}$. As a consequence, it should be possible to
further increase the coherence time by increasing the detuning of
the trapping laser.

\section{Simultaneous Ramsey measurements}\label{simultaneous}
Two-dimensional registers of qubits in arrays of dipole traps
based on microoptical elements are central for our implementation
of quantum information processing with cold neutral atoms. It is
essential to demonstrate the applicability of the techniques
described in the previous sections to atoms trapped in dipole trap
arrays. This includes the demonstration of parallel qubit
manipulation and a scheme for simultaneous but also site-specific
detection of the outcome of the respective operations.
For this purpose, we performed a simultaneous Ramsey experiment on
16 atom samples trapped in a 4 x 4 array of dipole traps with a
readout of the results based on position resolved imaging. The
dipole traps are based on a diffractive array of microlenses with
a total of 50 x 50 lenses. The lenses are separated by
\unit{125}{\micro\metre} and have a nominal focal length of
\unit{625}{\micro\metre}. The array is illuminated by a linear
polarized titanium sapphire laser beam with a power of
\unit{130}{\milli\watt} and wavelength of \unit{800}{\nano\metre}.
Due to the limited diffraction efficiency, about $40\%$ of the
incident light contribute to dipole trapping. The focal plane of
the array is imaged onto the MOT by a telescope which consists of
an achromat with a focal length of \unit{80}{\milli\metre} and the
lens system with a working distance of \unit{36}{\milli\metre}.
The trap array is resized to a distance between foci of
\unit{54}{\micro\metre}. We measured the waists of each trap at
the position of the MOT to \unit{1.7}{\micro\metre}. Due to the
Gaussian shape of incident beam illuminating the lens array, the
depths of the dipole traps vary from outer to inner traps with
$k_{B}\cdot\unit{600}{\micro\kelvin}$ for the weakest
and $k_{B}\cdot\unit{1.2}{\milli\kelvin}$ for the deepest traps.\\
\begin{figure}[b]
    \begin{center}
        \resizebox{0.4\textwidth}{!}{%
        \includegraphics{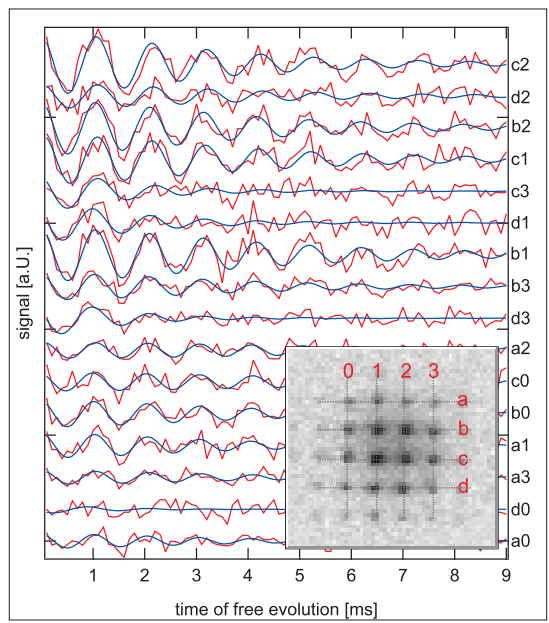}}
    \end{center}
    \caption{Simultaneous Ramsey measurements in 16 different dipole traps. The associated position of the
    traps is shown in the inset. The inset is a fluorescence image of the atoms trapped in the
    two-dimensional trap array.}
    \label{fig:Ramseys}       
\end{figure}
After loading the atoms from the MOT, we end up with more than 16
filled traps with around $500$ atoms in the central trap. The
population of the outer traps is reduced by more than one order of
magnitude because of the lower trap depth and the inferior overlap
with the optical molasses. The temperature of the atom ensembles
in the traps was determined to be less than
$\unit{50}{\micro\kelvin}$. The previous mentioned techniques for
coherent manipulation are simply applicable to the trap array by
illuminating all traps with Raman laser light simultaneously. Also
the detection scheme is fully adoptable to trap registers: we
perform the readout of the final result by taking spatially
resolved images of fluorescence emitted by the atoms (see inset in
Fig. \ref{fig:Ramseys}). The images are analyzed by integrating
the fluorescence around each known position of an atom sample and
thus achieving a site-specific determination of the population of
the qubit states at each site. Figure \ref{fig:Ramseys} shows
simultaneous Ramsey measurements for 16 register sites. Clearly
visible are Ramsey oscillations in almost all of the atom samples.
The variation in signal strength reflects the variation in atom
number. This proves that parallel coherent manipulation has been
achieved. Together with the site-specific addressability as
presented in \cite{Dumke1} this is a clear demonstration of the
scalability of our system and the techniques used for preparing,
manipulating, and readout of the qubit states.
\section{Conclusion and Outlook}
\label{conclusions} With this work, we have systematically studied
the coherent manipulation of atomic qubits in optical
micropotentials based on individual dipole traps and trap arrays.
By applying a range of techniques, like Rabi, Ramsey, and
spin-echo techniques, we could investigate dephasing and
decoherence of atomic qubits in dipole traps. We could show that
our approach for quantum information processing based on trapping
atomic qubits in arrays of dipole traps allows for a simultaneous
coherent manipulation of a large number of atomic qubits. This
presents an important step towards the realization of a scalable
quantum processor with atomic qubits. With the results described
in this paper ($\pi/2$ times of approx.
$\unit{250}{\micro\second}$ and a coherence time of
$\unit{68}{\milli\second}$) we already can perform more than 100
single qubit gate operations within the coherence time. In
addition, this value can be significantly increased: so far
decoherence is given by the inverse of the spontaneous scattering
rate of the trapping light. This rate can be reduced by increasing
the detuning of the trapping light. For traps based on the light
of an Nd:YAG laser (or similar) at around 1064 nm, decoherence
times in excess of $\unit{1}{\second}$ are achievable for
comparable trap depths. However, for the measurements presented
above the light of the Raman laser was distributed over a size of
about $\unit{1}{\centi\meter^2}$ in order to illuminate all
trapping sites homogenously. As the next step, this light can be
concentrated onto the known sites of the atom samples. This will
allow increasing of the intensity of the Raman light by more than
a factor of 100 thus reducing the $\pi/2$ time to a value well
below $\unit{10}{\micro\second}$. With these two measures the
number of single qubit gate operations can be increased to a value
of more than $10^5$ during the coherence time, bringing the system
into the parameter range necessary for complex quantum gate
schemes and the implementation of quantum error correction codes.\\
One of our current projects is the demonstration of a detection
sensitivity down to one atom per register site. Estimates based on
the known parameters of the detection scheme and the
specifications of the CCD camera together with initial experiments
on the current detection limit allow us to expect that single atom
detection is possible in the setup described in this work. This
will allow opening of work in an additional direction: a
two-dimensional array of well separated trapped single atoms can
serve as an excellent basis for a new scheme of atomic clocks. In
recent work, the applicability of atoms in optical lattices
\cite{Katori1} and microchip traps \cite{Reichel1} as a frequency
reference has been discussed. For optical traps, using trapping
light at a so called 'magic' wavelength (as is possible for
strontium) minimizes the effect of differential light shifts on
the frequency of the clock transition. For $^{85}$Rb, it was shown
that an additional light field tuned between the two hyperfine
ground states can also be used to suppress the differential light
shift \cite{Kaplan1}. Combining these results with a trapping
geometry based on arrays of microlenses would allow creation of a
large register of single trapped atoms in a configuration where
each atom can be spatially resolved. This allows treatment of each
atom as an individual atomic clock which can be separately
observed but at the same time gives a large number of individual
clocks for increasing the signal-to-noise ratio or even making use
of the superior performance of a massively entangled atom system.
\section{Acknowledgements}
\label{acknowledgements} This work has been financially supported
in part by the following funding organizations: DFG
(Schwerpunktprogramm 'Quanteninformationsverarbeitung'), European
Commission (IP 'ACQP', IP 'SCALA', RTN 'Atom Chips'),
NIST/NSA/ARDA/DTO (QCCM Project 'Neutral Atom Quantum Computing
with Optical Control'), and Land Hessen ('Innovationsbudget
Quanteninformationverarbeitung').
\\
\\
\\
\\
%
%
%
%

\begin{thebibliography}{}
%
%
\bibitem{QIP_Overview}
M.A. Nielsen, I.L. Chuang, \textit{Quantum Computation and Quantum
Information} (Cambridge University Press, Cambridge, 2000); T.
Beth, G. Leuchs (eds.) \textit{Quantum Information Processing}
(Wiley-VCH, Weinheim 2005)

\bibitem{Atomchips_Overview}
For an overview see: R. Folman, P. Kr\"uger, J. Schmiedmayer, J.
Denschlag, C. Henkel, Adv. At. Mol. Opt. Phys. \textbf{48,} (2002)
263

\bibitem{Grangier1}
N. Schlosser, G. Reymond, I. Protsenko, P. Grangier, Nature
(London) \textbf{411,} (2001) 1024

\bibitem{Optical_Lattices}
P.S. Jessen, I.H. Deutsch, Adv. At. Mol. Opt. Phys. \textbf{37,}
(1996) 95; I. Deutsch, P.S. Jessen, Phys. Rev. A \textbf{57,}
(1998) 1972; G. Grynberg, C. Robbilliard, Phys. Rep. \textbf{355},
(2001) 355

\bibitem{Birkl1}
G. Birkl, F.B.J. Buchkremer, R. Dumke, W. Ertmer, Opt. Comm.
\textbf{191,} (2001) 67

\bibitem{Dumke1}
R. Dumke, M. Volk, T. M\"uther, F.B.J. Buchkremer, G. Birkl, W.
Ertmer, Phys. Rev. Lett. \textbf{89,} (2002) 097903


\bibitem{Klempt1}
C. Klempt, T. van Zoest, T. Henninger, O. Topic, E. Rasel, W.
Ertmer, J. Arlt, Phys. Rev. A \textbf{73,} (2006) 013410


\bibitem{Davidson1}
N. Davidson, H.J. Lee, C.S. Adams, M. Kasevich, S. Chu, Phys. Rev.
Lett. \textbf{74,} (1995) 1311

\bibitem{Ozeri1}
R. Ozeri, L. Khaykovich, N. Davidson, Phys. Rev. A \textbf{59,}
(1999) R1750

\bibitem{Kuhr1}
S. Kuhr, W. Alt, D. Schrader, I. Dotsenko, Y. Miroshnychenko, A.
Rauschenbeutel, D. Meschede, Phys. Rev. A \textbf{72,} (2005)
023406

\bibitem{Bloch1}
F. Bloch, Phys. Rev. \textbf{70,} (1946) 460

\bibitem{Ramsey1}
N.F. Ramsey, Phys. Rev. \textbf{78,} (1950) 695



\bibitem{Hahn1}
E.L. Hahn, Phys. Rev. \textbf{80,} (1950) 580

\bibitem{Kurnit1}
N.A. Kurnit, I. D. Abella, S. R. Hartmann, Phys. Rev. Lett.
\textbf{13,} (1964) 567

%

\bibitem{Buchkremer1}
F.B.J. Buchkremer, R. Dumke, H. Levsen, G. Birkl, W. Ertmer, Phys.
Rev. Lett. \textbf{85,} (2000) 3121

\bibitem{Andersen1}
M.F. Andersen, A. Kaplan, N. Davidson, Phys. Rev. Lett.
\textbf{90,} (2003) 023001

\bibitem{Katori1}
H. Katori, M. Takamoto, V.G. Pal'chikov, D. Ovsiannikov, Phys.
Rev. Lett. \textbf{91,} (2003) 173005

\bibitem{Reichel1}
P. Treutlein, P. Hommelhoff, T. Steinmetz, T.W. H\"ansch, J.
Reichel, Phys. Rev. Lett. \textbf{92,} (2004) 203005

\bibitem{Kaplan1}
A. Kaplan, M.F. Andersen, N. Davidson, Phys. Rev. A \textbf{66,}
(2002) 045401

%
\end{thebibliography}
%

\end{document}